\documentclass[journal]{IEEEtran}
\usepackage{multirow}
\usepackage{booktabs}
\usepackage{cite}
\usepackage{bm}
\usepackage{color}
\ifCLASSINFOpdf
   \usepackage[pdftex]{graphicx}
   \graphicspath{{../pdf/}{../jpeg/}}
   \DeclareGraphicsExtensions{.pdf,.jpeg,.png}
\else
   \usepackage[dvips]{graphicx}
   \graphicspath{{../eps/}}
   \DeclareGraphicsExtensions{.eps}
\fi
\begin{document}

\title{Dynamic Modulation of Electromagnetically Induced Transparency Metamaterials through Mode Coupling and Stretchable Design} 

\author{Sihong Chen,~\IEEEmembership{Member, IEEE,}
	Taisong Pan,~\IEEEmembership{}
	Zhengcheng Mou,~\IEEEmembership{}
	Bing-Zhong Wang,~\IEEEmembership{Senior Member, IEEE,}
	and Yuan Lin~\IEEEmembership{}
	\thanks{Sihong Chen is with the School of Materials and Energy, School of Physics, University of Electronic Science and Technology of China, Chengdu {\rm 610054}, China, and also with Huawei Technologies, Chengdu {\rm 610041}, China (e-mail:Sihongchen@outlook.com or chensihong1@huawei.com).}
	
	\thanks{Taisong Pan is with the School of Materials and Energy, University of Electronic Science and Technology of China, Chengdu {\rm 610054}, China (e-mail:tspan@uestc.edu.cn).}
	
	\thanks{Zhengcheng Mou is with the School of Statistics, Southwestern University of Finance and Economics, Chengdu {\rm 611130}, China (e-mail:mzc971123@foxmail.com).}	
	
	\thanks{Bing-Zhong Wang is with the Institute of Applied Physics, University of Electronic Science and Technology of China, Chengdu {\rm 611731}, China (e-mail:bzwang@uestc.edu.cn).}
	
	\thanks{Yuan Lin is with the School of Materials and Energy, Medico-Engineering Cooperation on Applied Medicine Research Center, University of Electronic Science and Technology of China, Chengdu {\rm 610054}, China (e-mail:linyuan@uestc.edu.cn).}
	\thanks{Corresponding author: Taisong Pan}
}

\maketitle
\begin{abstract}
The active control of electromagnetically induced transparency (EIT) metamaterials (MM) has the potential to revolutionize communication networks without relying on quantum technology. However, current reconfigurable systems offer limited flexibility and have high fabrication costs and difficulties. In this study, we examine a classical EIT metamaterial and discover a novel modulation mechanism that leverages mode coupling to dynamically adjust the bandwidth and group delay of the EIT MM. This mechanism is verified through analyses of the electric field and surface charge density distributions. Additionally, a robust coupled Lorentz oscillator model is used to explain the coupling mechanism, with results that are in good agreement with simulations and experiments. To capitalize on this mechanism, we propose a block-definition approach where the MM is divided into stretchable sections, allowing for dynamic modulation of the bandwidth and group delay by stretching the EIT MM. Furthermore, the fabrication process is highly compatible with traditional flexible printed circuit board techniques. Our block-definition EIT MM offers unprecedented tunability and flexibility, requiring no complex components or specialized materials, making it a promising candidate for tunable slow-wave devices and other reconfigurable microwave applications.
\end{abstract}
\section{Introduction}
Metamaterials (MMs) provide a diverse and incredible way to manipulate electromagnetic waves, making them widely used in sensing \cite{doi:10.1063/1.4906109,7393446}, electromagnetic cloaking \cite{Schurig977}, perfect absorption\cite{physRevLett.100.207402,doi:10.1002/adma.201200674,Alaee12,doi:10.1063/1.3608246}, hologram imaging \cite{zheng2015metasurface} and polarization conversion \cite{zhao2012twisted}. The phenomenon of electromagnetic induced transparency (EIT) is a quantum interference effect that creates a window of transmission within the absorption band of a three-level atomic system\cite{PhysRevLett.64.1107}. This leads to extreme modifications in dispersion characteristics and can have many potential applications, such as slow light\cite{hau1999light,lukin2001controlling,doi:10.1002/lpor.201100021} and enhanced non-linear effects\cite{PhysRevLett.99.123603,Schmidt96}. However, achieving this effect through conventional quantum EIT methods requires harsh experimental conditions. To overcome this limitation, the use of MMs has been proposed to mimic the original quantum EIT effect\cite{yahiaoui2017active,PhysRevLett.101.253903,PhysRevLett.101.047401,PhysRevB.79.085111, PhysRevLett.106.107403,liu2009plasmonic,hao2008symmetry,PhysRevB.94.161103,xiao2018active,PhysRevA.92.023818,doi:10.1119/1.1412644}.

As a result, various MM-based EIT applications have been reported\cite{yan2019terahertz,yang2019electromagnetically,chen2021analogue,chen2020noncontact,zhang2021highly,ning2019wideband,zhang2020active,han2020polarization,guo2018enhancement,litt2018hybrid,zhu2022electromagnetically}, with slow-light devices being one of the most significant\cite{ning2019wideband,zhang2020active,han2020polarization}. The delay-bandwidth product (DBP) is a critical metric that measures the performance of slow light, as it is the product of the maximum group delay and full width at half maximum (FWHM) bandwidth\cite{pitchappa2016active,li2015electromagnetically}. To achieve good results in slow light, it is important to have a large group delay time and bandwidth for EIT MMs. However, the EIT effects based on MMs have a very narrow transmission window due to their intrinsic resonant characteristics\cite{yahiaoui2017active}.

Although there have been some reports demonstrating broadband EIT effects in terahertz and optical regime\cite{su2015broadband,yahiaoui2017active,wu2011broadband}, the quest for new designs to achieve a transparent window over a broad spectral range is still urgent and remains a challenging issue, especially in the low-frequency regions. Additionally, in some situations, real-time reconfigurability is needed in response to rapidly varying operational scenarios, but conventional EIT MMs cannot meet these requirements. Their fixed properties seriously hamper the developments and practical applications of EIT effects. Some efforts have been made to modify the EIT effect in MMs through hybrid EIT MMs\cite{gu2012active,chen2008experimental,doi:10.1002/adma.201603355,xu2016frequency,wang2017hybrid,dicken2009frequency,cao2013plasmon} and MEMS reconfigurable EIT MMs\cite{doi:10.1063/1.4969061,doi:10.1063/1.4943974,doi:10.1002/adom.201500676}. However, in these strategies, the active control of EIT effect is basically achieved by integrating adjustable materials and components or adopting tedious preparation processes, which increases costs and hinders their applications. Also, these strategies are not suitable for the applications in microwave applications, as they require large areas of adjustable materials and components, and the fabrication of large-area reconfigurable MEMS-based MMs is a challenge.

Recently, stretchable devices, implemented on polymers with low surface energy, can attach to arbitrary surfaces in a conformal manner\cite{badloe2021tunable,qi2021stretchable}. This flexibility allows for the integration of passive or functional devices with curved surfaces and packaging materials, unlike rigid platforms. This new development in MMs opens up the possibility for reconfigurable or multi-functional devices.

Thus, in this paper, we introduce a new modulation mechanism that utilizes mode coupling tuning to achieve reconfigurable bandwidth and group delay in EIT analogue. This new mechanism was inspired by stretchable MMs. After a comprehensive analysis of a classical EIT structure, we uncovered the new modulation mechanism and demonstrated it with the help of a robust coupled Lorentz oscillator mode, which showed good alignment with simulations and confirmed by the analysis of electric field and surface charge density distributions. Prototype devices were also fabricated and tested, with results consistent with simulations. To realize dynamic modulation of the frequency response and group delay, we proposed defining the devices into stretchable blocks using transfer printing and laser cutting techniques. Both simulations and experiments indicate that the group delay and bandwidth can be adjusted in a continuous manner by stretching the device. The proposed EIT MM fabrication process is compatible with traditional flexible PCB techniques and does not require complex components or specialized materials. This approach has great potential for tunable slow-wave devices and other microwave tunable devices, offering on-the-fly tunability and high flexibility for various specifications such as microwave buffering and active filtering.

\section{Design Concept of the EIT MMs}
\begin{figure}[!h]
	\centering
	\includegraphics[scale=0.4]{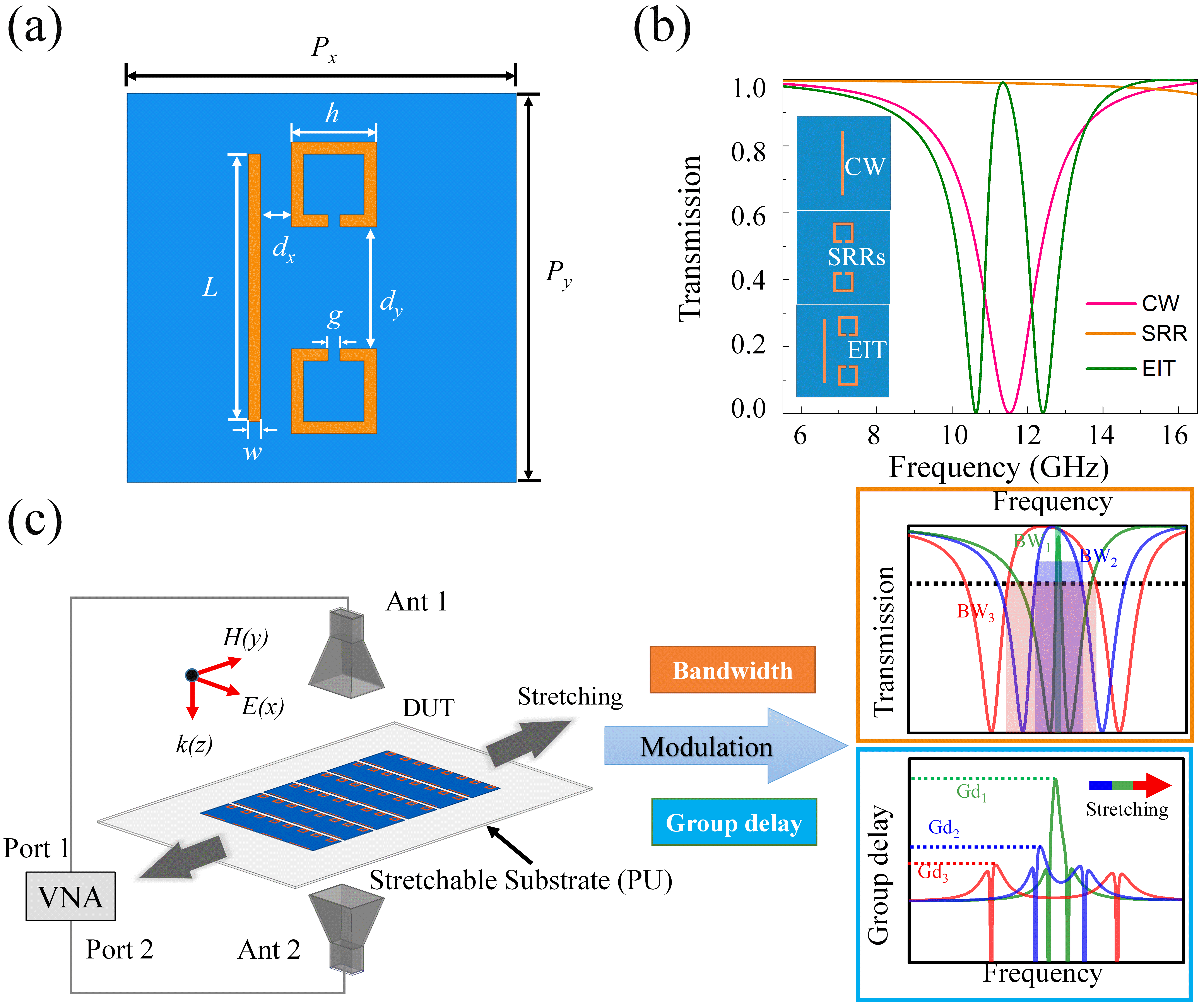}
	\caption{(a) The geometrical illustration of the proposed EIT MM. The relevant geometrical parameters are as follows: $P_x$ = $P_y$ = 16 $mm$, $d_x$ = 2.0 $mm$, $h$ = 3.5 $mm$, $g$ = 0.5 $mm$, $d_y$ = 5 $mm$, $w$ = 0.5 $mm$ and $L$ = 11 $mm$. (b) Simulated transmission response of the CW, SRRs and EIT MM. Insets show the geometries of the CW, the SRRs and the EIT MM, respectively. (c) Experimental setup of the EIT MM and corresponding demonstration of its modulation effect on bandwidth and group delay with stretching.}
	\label{fig_sckematic}
\end{figure}

The schematic illustration of the analyzed MMs is presented in Fig. \ref{fig_sckematic}(a), which employed a kind of classical structure to produce the EIT-like effect, similar to the reference reported by Jianqiang Gu \textit{et al}.\cite{gu2012active}. The unit cell of 0.018 mm-thick copper (Cu)-based structure consists of a cut wire (CW) and two split-ring resonators (SRRs) fabricated on flexible polyimide substrate (relative permittivity: 3.5, loss tangent: 0.008 and thickness: 0.1 $mm$). The lattice constant of array is $P_x$ = $P_y$ = 16 $mm$ and the length and width of the CW is $L$ = 11 $mm $ and $w$ = 0.5 $mm$, respectively. We use $d_x$ to describe the coupling distance between the SRRs and CW with a initial value $d_x$ = 2.0 $mm$. The SRRs are $h$ = 3.5 $mm$ in side length and $g$ = 0.5 $mm$ in split gap and the $d_y$ = 5 $mm$ represents the distance between the SRRs. To clarify the EIT analogue of the MMs in microwave region, three sets of arrays are simulated with the unit cell composed of a CW, a pair of SRRs and both of them, respectively, and the corresponding transmission responses are shown in Fig. \ref{fig_sckematic}(b). It can be seen that the CW exhibits a typical electric resonance at 11.5 $GHz$ \cite{li2015low,yahiaoui2017active}, which has a broad line width because of the strong radiation losses under the normal incident plane wave with the electric field parallel to it, whereas the SRRs, supporting an LC resonance mode with a narrow line width at the same frequency, are inactive due to the incident electric field perpendicular to the structure. When both of them are arranged into the unit cell, the field coupling will occur between the CW and the SRR-pair, in which the CW that exhibits electric resonance and the SRRs that excite the LC resonance serve as the bright and dark mode, respectively. The destructive interference between the electric resonance and LC resonance leads to appearance of the EIT-like behavior. Fig. \ref{fig_sckematic}(c) demonstrates the design concept of dynamic modulation of frequency response and group delay of the device. Vector Network Analyzer (VNA) is used to test the transmission response of the proposed devices, to which two ridge antennas are connected to the VNA and placed in the line of sight to each other with the fabricated device placed in the middle. The frequency response and group delay of the device can be continuously modulated by controlling the coupling distance of stretchable EIT MMs with a customized stretching holder.

\begin{figure*}[!h]
	\centering
	\includegraphics[width=0.72\textheight]{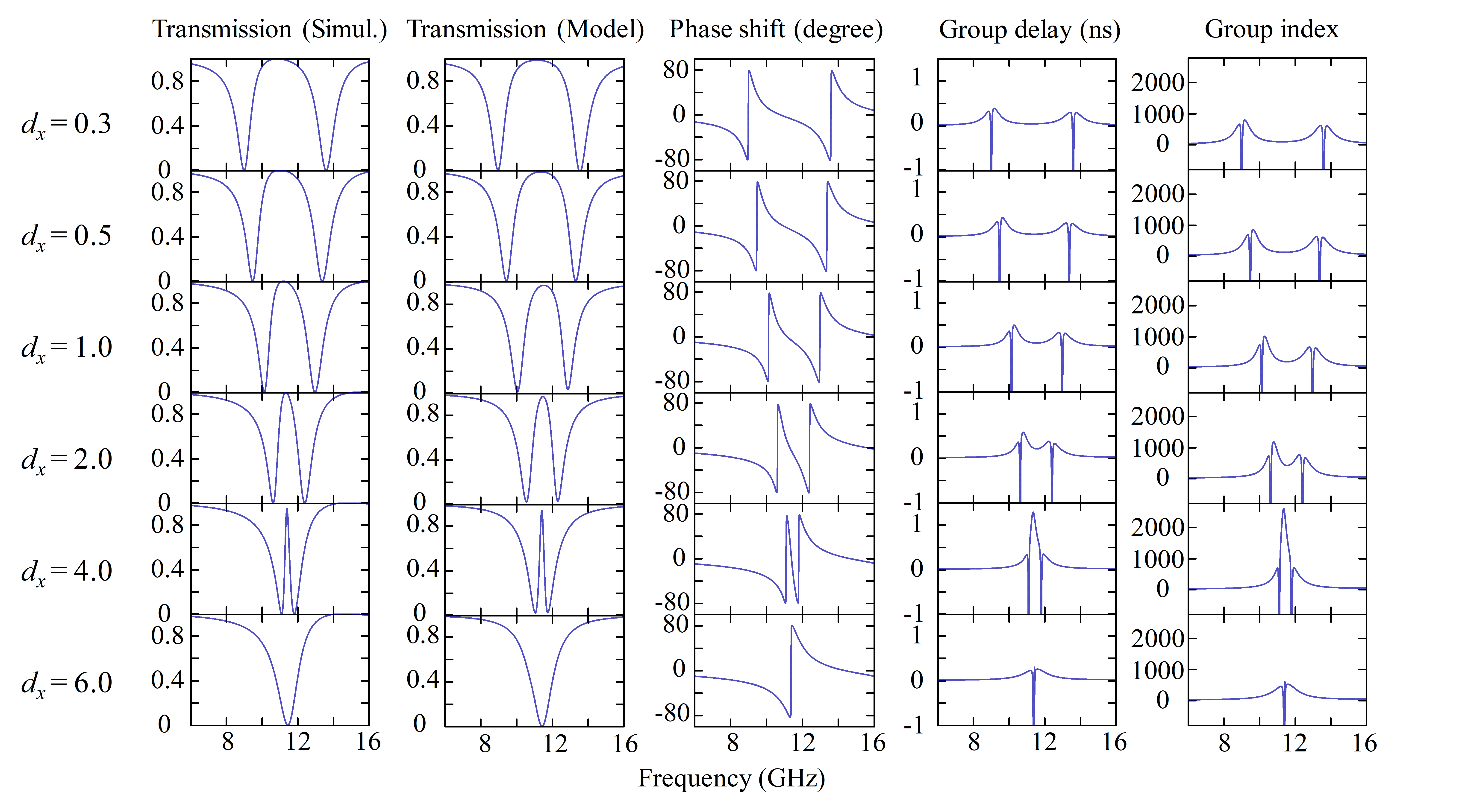}
	\caption{Spectra of the transmission amplitude (simulation and coupled Lorentz oscillator model) and phase, group delay and group index of EIT MM with different coupling distance.}
	\label{fig_parametic_sweep}
\end{figure*}
\section{Analyses of the EIT MMs}
To verify the feasibility of the design concept, we did detailed analyses on the proposed structures. The full-wave numerical simulation software CST Microwave Studio is employed to analyze the spectral response, as shown in Fig. \ref{fig_parametic_sweep}. In these calculations, open boundary conditions are set along the electromagnetic wave propagating direction (z-direction), and periodic boundary conditions are applied to mimic the function of a 2D infinite condition. As discussed in previously reports\cite{tassin2012electromagnetically}, the coupling mechanism between the SRRs and CW plays a key role in controlling the EIT effect. Thus, we firstly studied the transmission characteristics with different coupling distance ($d_x$), as demonstrated in Fig. \ref{fig_parametic_sweep}. By tuning the coupling distance ($d_x$), we plot the transmission amplitude and phase, group delay and group index for a set of MMs with different $d_x$. As we can see that there is a strong correlation between the bandwidth and group delay of EIT effect and coupling distance, in which the bandwidth of the EIT narrows with the increase of $d_x$, along with a gradually increase of the max group delay and group index. Notably, the disappearance of the transmission window is observed at the $d_x$ of 6 $mm$, with the same situation of group delay. 

To explore the physical mechanism of this behavior, we employed the coupled Lorentz oscillator model to quantitatively explain the near field coupling between the bright and dark mode with the change of $d_x$ \cite{PhysRevLett.101.047401,tassin2012electromagnetically}. Similar to quantum EIT in the three-level system, we employed a ground state $|0 \rangle$ and two excited states $|1 \rangle$ and $|2 \rangle $ to illustrate the model, of which the transitions $|0 \rangle-|1 \rangle $ and $ |0 \rangle-|2 \rangle $ represent the dipole-allowed transition and dipole-forbidden transition, corresponding to the typical dipole bright mode resonance in the CW and the LC dark mode resonance in the SRRs, respectively. Then, the transition between $|1 \rangle$ and $|2 \rangle$ illustrates the coupling between the CW and the SRRs. Because there exist two possible excitation pathways of $|0 \rangle-|1 \rangle $ and $|0 \rangle-|1 \rangle-|2 \rangle-|1 \rangle$, introducing destructive interference, which leads to strong suppression of loss and large dispersion within a narrow frequency band. The interference in the MMs can be analytically described by the coupled differential equations
\cite{yahiaoui2017active,han2016tunable,tian2017low},

\begin{equation}
	\centering
	\ddot{x_1}(t)+\gamma_1\dot{x_1}(t)+\omega_0^2 x_1(t)+\Omega x_2 (t) = gE_0(t)
	\label{eq1}
\end{equation}
\begin{equation}
	\centering
	\ddot{x_2}(t)+\gamma_{2}\dot{x_2}(t)+(\omega_0+\delta)^2x_2(t)+\Omega x_1(t) = 0
	\label{eq2}.
\end{equation}

Where $x_1$ and $x_2$ represent the amplitudes and $\gamma_1$ and $\gamma_2$ describe the radiating rate of bright and dark mode, respectively. Then the parameter $g$ describes the coupling strength between the bright mode and the incident electromagnetic field and the coupling coefficient between the resonators is represented by $\Omega$. The difference between resonance frequency of intrinsic oscillators and transparency frequency is described by the detuning factor $\delta$. By solving the coupled equations \ref{eq1} and \ref{eq2} using the displacement vector $\bm{x_n}(t)= x_n e^{i\omega t}(n = 1,2)$  and the approximation $\omega_1^2-\omega^{2}\approx -2\omega_1(\omega-\omega_1) $\cite{liu2009plasmonic}, we can get the transmission coefficient based on the relation of $T=1-R$\cite{PhysRevLett.102.053901}. 

\begin{equation}\small
	\centering
	T = 1 - \rm{Re} \frac{ig^2(\omega-\omega_0-\delta+i\gamma_2/2)}{(\omega-\omega_0+i\gamma_1/2)(\omega-\omega_0-\delta+i\gamma_2/2)-\Omega^2/4}
	\label{eq3}.
\end{equation}

\begin{figure}[!h]
	\centering
	\includegraphics[scale=0.32]{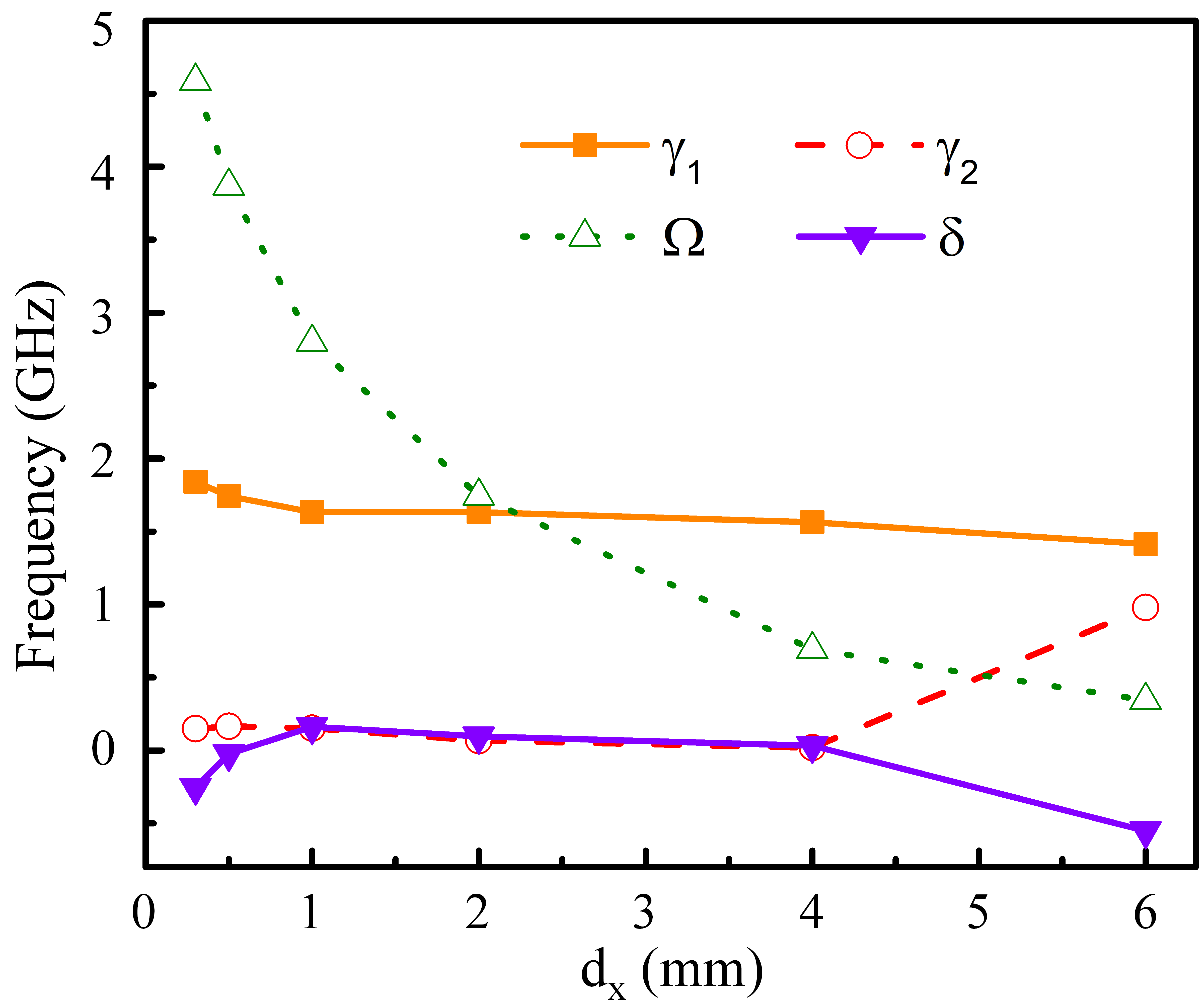}
	\caption{The fitting parameters $\gamma_1$, $\gamma_2$, $\delta$ and $\Omega$ as a function of coupling distance $d_x$.}
	\label{fig_fiting_parameter}
\end{figure}

The analytical modeled transmission data using the equation \ref{eq3} are presented in Fig. \ref{fig_parametic_sweep} (Transmission Model), which exhibits good agreements with the simulations. The corresponding fitting parameters are listed in Table \ref{EIT_LCM_parameter} and plotted in Fig. \ref{fig_fiting_parameter}. It is observed that $\gamma_1$, $\gamma_2$ and $\delta$ do not change significantly with the increase of $d_x$, whereas the coupling coefficient ($\Omega$) decreases markedly, which indicates the coupled strength of the EIT MMs can be significantly tuned by changing the coupled distance $d_x$. Finally, with the increase of coupled distance, the coupling coefficient $\Omega$ becomes too small to excite the dark mode, which eventually results in the disappearance of the EIT transmission window.

\begin{table}[!h]
	\caption{The fitting parameters of coupled Lorentz oscillator model.}
	\label{EIT_LCM_parameter}
	\centering
	\begin{tabular}{|c|c|c|c|c|c|}
		\hline
		$d_x$ (mm) & $\gamma_1$ (GHz) & $\gamma_2$ (GHz) & $\Omega$ (GHz)& $\delta$ (GHz) & $g$ \\
		\hline
		0.3 & 1.841  & 0.1488 & 4.587 & -0.255 & 1.000   \\
		\hline
		0.5 & 1.743 & 0.165 & 3.868 & -0.0232 & 0.9753 \\
		\hline
		1 & 1.633  &  0.1515 & 2.797 & 0.1611 & 0.9324 \\
		\hline
		2& 1.632 & 0.0667  & 1.744 & 0.0965  & 0.9099 \\
		\hline
		4 & 1.563 & 0.0194 & 0.6914 & 0.03227 & 0.8797 \\
		\hline		
		6 & 1.414 & 0.979 & 0.3445 & -0.5516 & 0.8555 \\
		\hline
	\end{tabular}
\end{table}

\begin{figure}[!h]
	\centering
	\includegraphics[scale=0.4]{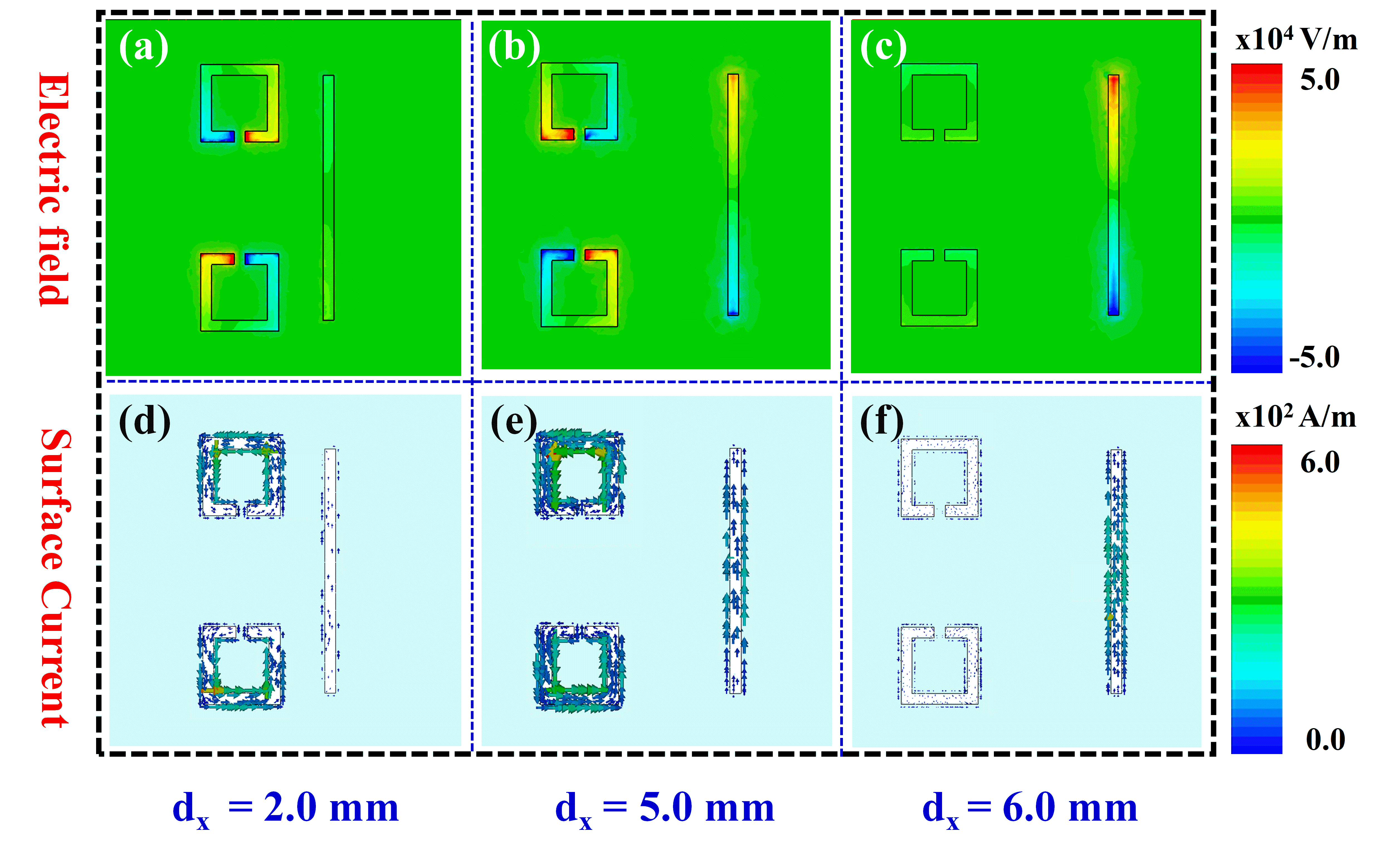}
	\caption{The simulated electric field and surface charge density distributions at the resonance frequency (11.5 $GHz$) of the EIT MMs with the coupling distance $d_x$ of 2.0 $mm$, 5.0 $mm$ and 6.0 $mm$, respectively.}
	\label{fig_Electric_field}
\end{figure}

Furthermore, to figure out the coupling behavior of the proposed device, the electric field and surface charge density distributions at the resonance frequency are also presented to illuminate mode coupling process. Fig. \ref{fig_Electric_field} shows the distributions at different coupling distances ($d_x$ = 2.0 $mm$, $d_x$ = 5.0 $mm$, and $d_x$ = 6.0 $mm$), corresponding to different coupling strength resulting in the pronounced EIT peak, EIT disappearance and only electric dipole resonance, respectively. As we can see in Fig. \ref{fig_Electric_field}(a), when $d_x$ is 2.0 $mm$, the electric field distributions are mainly concentrated around the SRRs' gap, whereas those in the CW are very weak. Meanwhile, in the analyzed structure, the incident waves cannot excite the LC resonance in the SRRs directly, so near-field coupling is the only way to excite LC resonance whose resonant frequency is the same with that of CW. When CW is excited by external waves, most of the energy is transferred to the SRRs by the electric field coupling, causing the suppression of the electric field in the CW, which is the characteristic of a typical EIT effect\cite{PhysRevLett.101.047401}. With the increase of $d_x$, the coupling strength gradually decreases and the electric field in the SRRs' gap is gradually suppressed, leading to a redistribution of the electric field in the EIT structures, as shown in Fig. \ref{fig_Electric_field}(b), where the CW and SRRs are both excited. Then, as the continuous increase of the coupling distance, the electric field in the SRRs' gap is completely suppressed and the electric field distribution is primarily concentrated in the CW, as show in Fig. \ref{fig_Electric_field}(c), which behaves as a typical electric dipole resonance. Similar behaviors can be obtained in the distributions of surface current. Specifically, in Fig. \ref{fig_Electric_field}(d), surface current is mainly concentrated on the SRRs in the strong coupling state ($d_x$ = 2 $mm$). Then, as coupling distance increases to 5.0 mm, the surface current is redistributed to have both the SRR and CW excited with strong current flows shown in Fig. \ref{fig_Electric_field}(e). With the further increase of coupling distance, the coupling disappears, so that the surface current is mostly focused on the CW, showing the typical characteristics of electric dipole resonance, as shown in Fig. \ref{fig_Electric_field}(f). Thus, the origin of modulation of EIT resonance is realized by tuning the coupling strength, namely, by changing the distance between the SRRs and CW.

\begin{figure}[!h]
	\centering
	\includegraphics[scale=0.4]{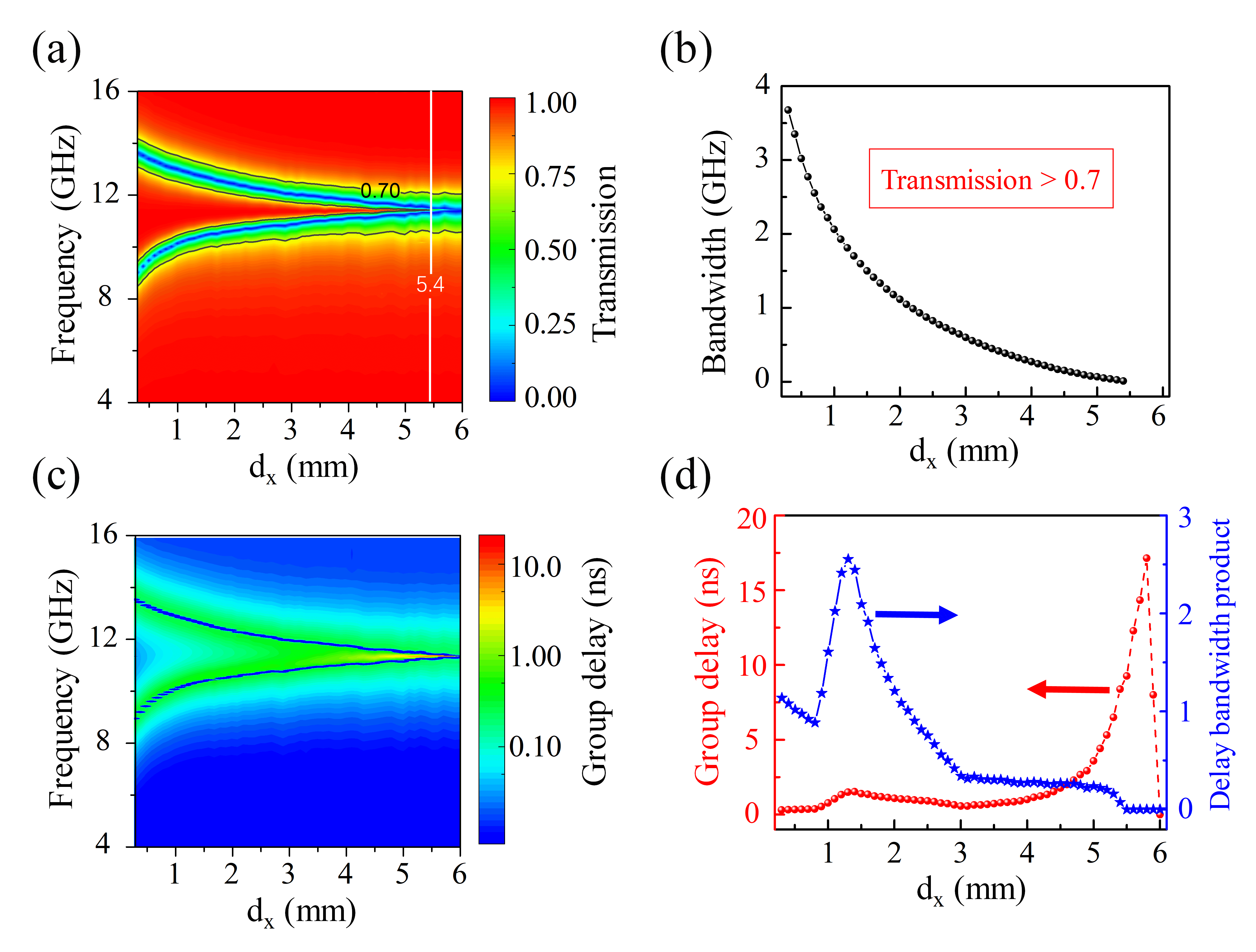}
	\caption{(a) The simulated 2D map of transmission response spectra as a function of frequency as well as coupling distance. (b) Transmission bandwidth with amplitude greater than 0.7. (c) 2D map of group delay as a function of frequency as well as coupling distance at. (d) The maximum group delay (red sphere-symbol line) and delay bandwidth product (blue star-symbol line) with the varied $d_x$. The step of $d_x$ is 0.1 $mm$}
	\label{fig_SRR_map}
\end{figure}

To obtain a deep insight into the evolution of the transmittance and group delay as a function of the $d_{x}$, more calculations are presented. We map the two-dimensional transmission response spectra as a function of frequency as well as coupling distance at the step of 0.1 $mm$, as shown in Fig. \ref{fig_SRR_map}(a). It shows that the transmission window gradually narrows with the increase of $d_{x}$, and the transmission amplitude is less than 0.7 at a distance greater than 5.4 $mm$. As is widely known, the prominent characteristics of the EIT phenomenon for slow-light devices are the product of the group delay time and bandwidth of transparency windows\cite{tsakmakidis2017breaking,zhao2018maximization}. Group delay ($\tau_g$) can be calculated by
\begin{equation}
	\centering
	\tau_g=-\frac{\partial{\phi}}{\partial{\omega}}
	\label{eq4},
\end{equation}
where $\phi$ is the transmission phase shift and $\omega$ is the angular frequency. The calculation of DBP can be calculated from the transmission FWHM bandwidth ($\Delta f$) and group delay as $DBP=\tau_g\times\Delta f$. Thus, we plotted the transmission bandwidth of the device with varied $d_{x}$ at a  transmission amplitude greater than 0.7 in Fig. \ref{fig_SRR_map}(b), showing a decreasing trend from 3.68 to 0.01 $GHz$ over the range from 0.3 to 5.4 $mm$. Then, a finer map of group delay as a function of frequency and $d_{x}$ is simulated in Fig. \ref{fig_SRR_map}(c) with the step of 0.1 $mm$ as well. Negative group delay appears at the side of resonance modes, while an obviously positive group delay occurs over the transparency window and a increasing trend is observed. Thus, we extract the maximum group delay and plot it as a function of $d_{x}$ into Fig. 5(d) marked with red sphere-symbol line. The achievement of maximum group delay is 17.14 $ns$ at the $d_{x}$ of 5.8 $mm$, but the transmission efficiency at the frequency is very small, thus being of little value for practical applications. Considering the transmission amplitude and group delay, the results of DBP with a transmission efficiency greater than 70\% are shown in Fig. 5(d) with blue star-symbol line. The maximum DBP achieves 2.55 at the $d_{x}$ of 1.3 $mm$.

\begin{figure}[!h]
	\centering
	\includegraphics[scale=0.38]{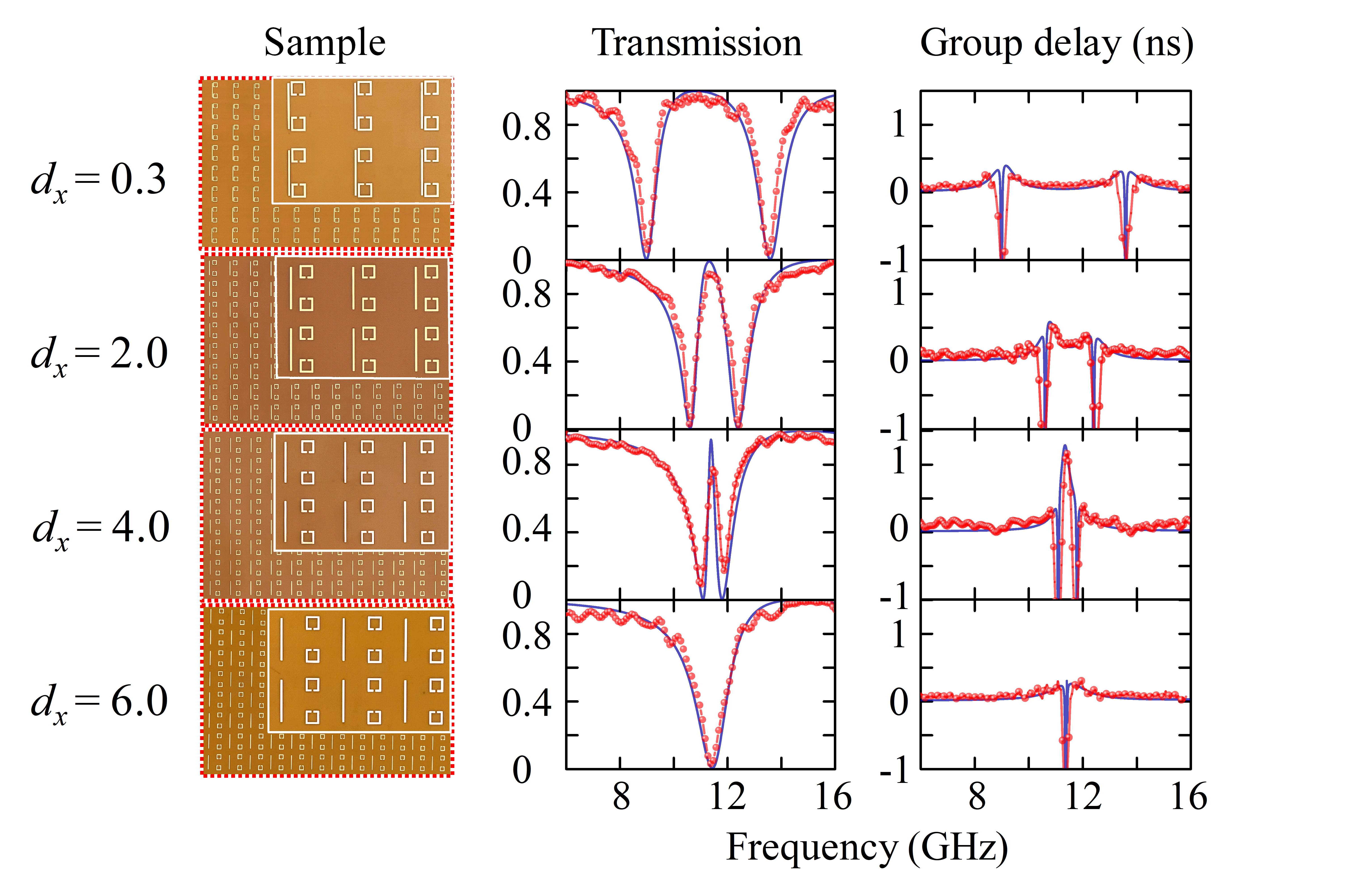}
	\caption{ The photographs of the fabricated EIT MM with different coupling distance $d_x$ and the simulated (blue solid line) and experimental (red sphere-symbol line) transmission and phase delay results.}
	\label{fig_SRR_test}
\end{figure}

To validate the simulated and modeled results, we fabricated four samples with different coupling distance, as presented in Fig. \ref{fig_SRR_test} and the simulated and experimental transmission and phase delay results are also demonstrated. As we can see, there are good agreements between the experimental results with the simulated ones, which proves that the simulation and theoretical analysis are reasonable. Meanwhile, the results manifest that the bandwidth and group delay of EIT MMs can be tuned by adjusting the coupling distance to realize a broadband or narrow band EIT effect, or to suppress EIT effect.

\section{Dynamic modulation of the EIT MMs}
\begin{figure}[!h]
	\centering
	\includegraphics[scale=0.42]{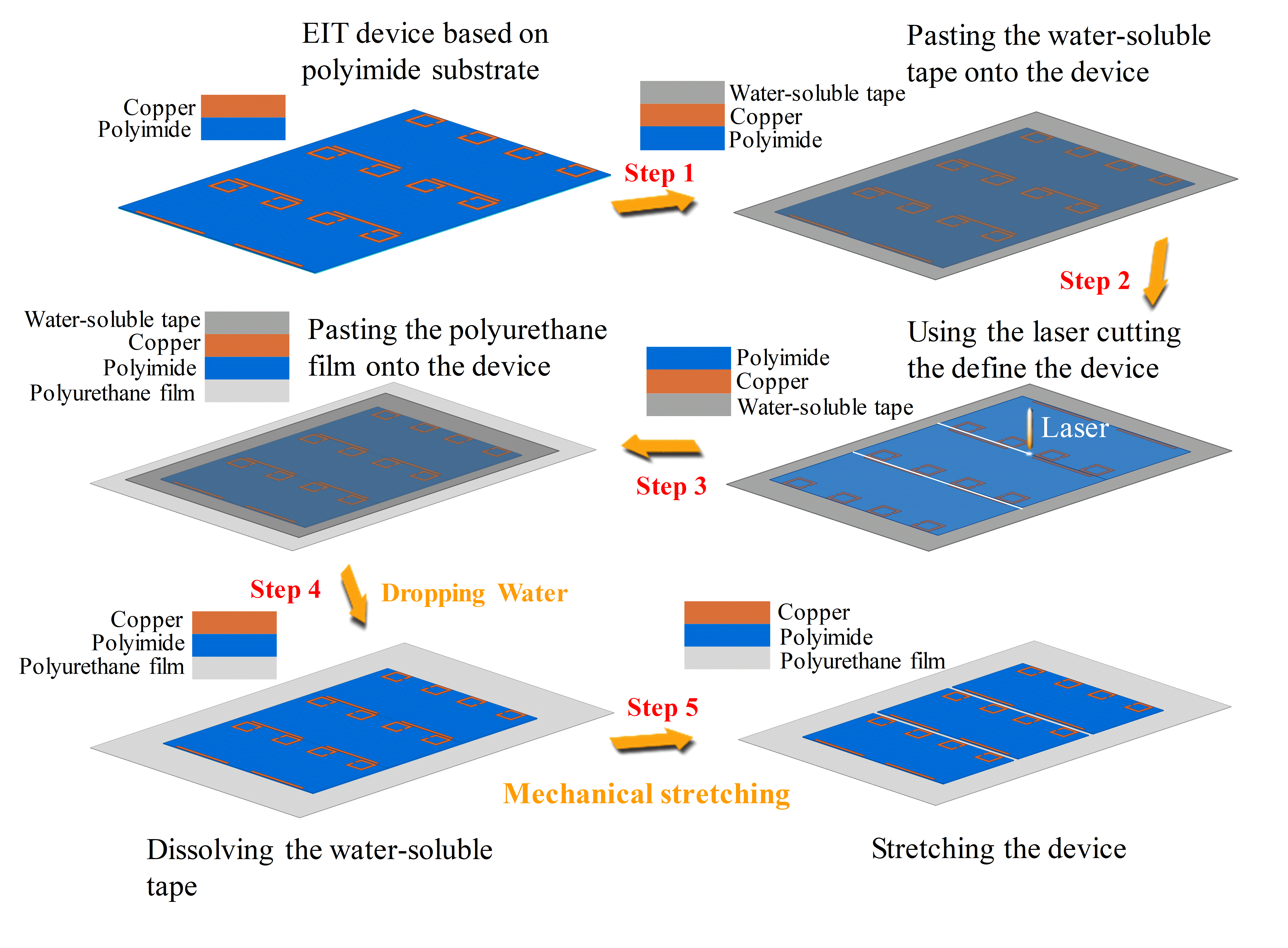}
	\caption{The detailed fabrication process of stretchable EIT device based on the method of laser-cutting and transfer printing.}
	\label{fig_process}
\end{figure}

Inspired by stretchable MMs and above analyses, we introduced an approach based on transfer printing and laser cutting techniques to achieve the dynamic modulation of the bandwidth and group delay of EIT MMs. The detailed fabrication processes are demonstrated in Fig. \ref{fig_process}. To fabricate the tunable EIT device, the process starts from pasting the water-soluble tape onto EIT device based on the polyimide substrate, as show in Fig. \ref{fig_process} of step 1. The EIT device is fabricated on the flexible polyimide substrate using the traditional PCB technology. Then, in the step 2, the laser-cutting is used to define the tunable blocks of the EIT device. In this step, in order to ensure the integrity of the device, we need to accurately control the laser power, so that it can cut through PI layer without cutting through the water-soluble tape layer below. After that, we should transfer the device with defined blocks onto the stretchable substrate layer (Polyurethane, PU) and use the water to remove the water-soluble tape, as shown in step 3 and 4. At this point, the tunable function devices are obtained and can realize the tunability by mechanical loading (step 5).

\begin{figure}[h!]
	\centering
	\includegraphics[scale=0.38]{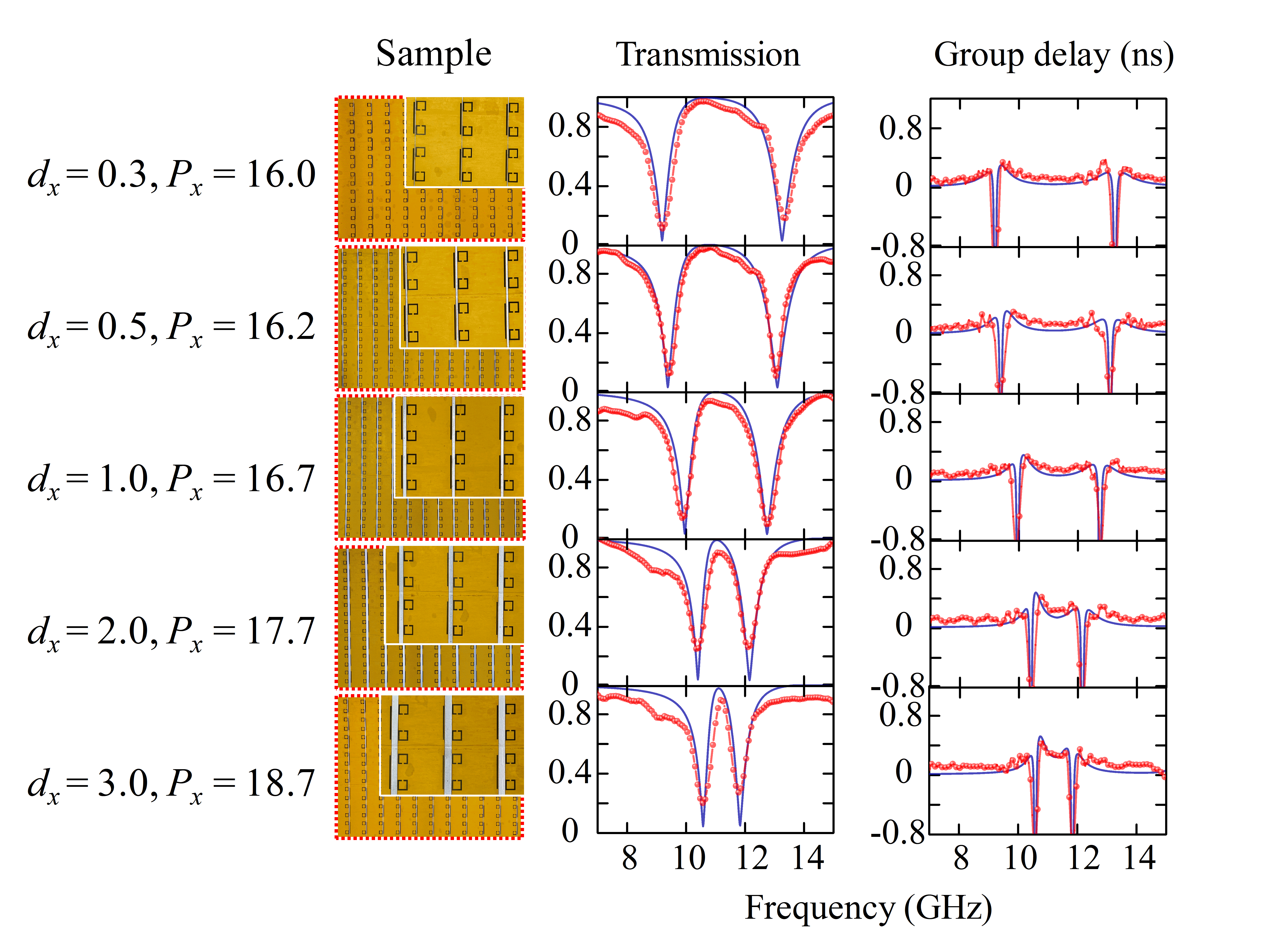}
	\caption{ The photographs of the tunable EIT MMs when stretched different distance and the simulated (blue solid line) and experimental (red sphere-symbol line) transmission and phase delay results.}
	\label{fig_SRR_stretching}
\end{figure}

Then we measured the stretchable device with the experiment setup shown in Fig. \ref{fig_sckematic}(c), and presented the photos of five states with different stretching distance and corresponding simulated and experimental transmission and phase delay results in Fig. \ref{fig_SRR_stretching}. The photos show a good control of the coupling distance in our measurement and the results also match well and demonstrate good modulation effects on the bandwidth and group delay by stretching the devices. 

\begin{figure}[h]
	\centering
	\includegraphics[scale=0.3]{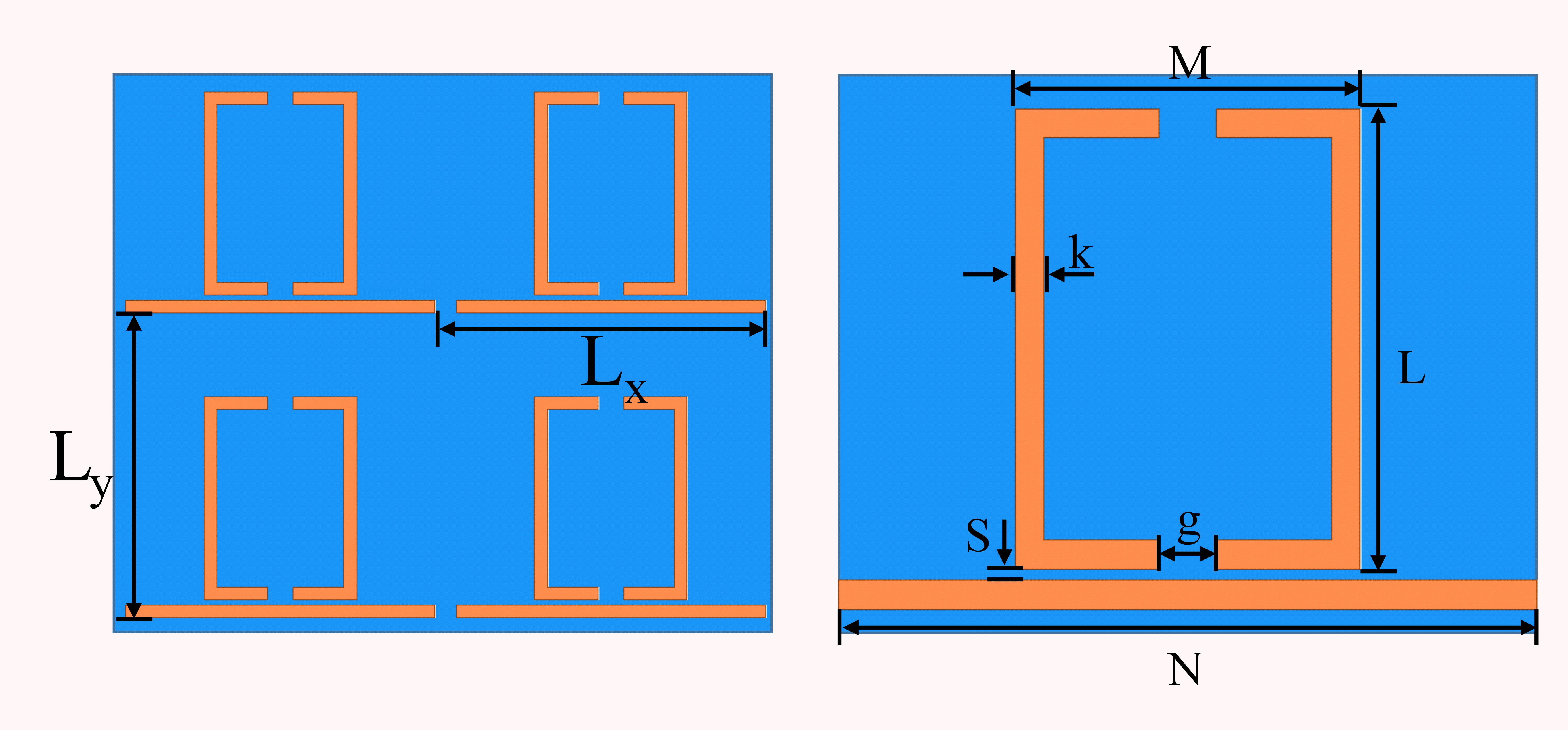}
	\caption{Geometric illustration of the optimized EIT MM (CW-DSR). The detailed values are set to $L_x$ = 5.0 $mm$, $L_y$ = 5.0 $mm$, $M$ = 6.0 $mm$, $N$ = 12.1 $mm$, $L$ = 8.0 $mm$, $k$ = 0.5 $mm$ and $g$ = 1.0 $mm$. The initial gap between the DSR and CW resonator is $S$ = 0.2 $mm$}
	\label{fig_DSR_skematic}
\end{figure}

However, we found that the modulation range has a certain decrease compared with that in Fig. \ref{fig_SRR_test} due to the extension of $P_x$, making it difficult to achieve that only electric dipole resonance is excited within the stretching range of substrate. Hence, for the realization of expected modulation effect, we optimized the original structure. The original SRRs are replaced with a double splits resonator (DSR) structure to form a basic unit of the CW-DSR structure and the details are shown in Fig. \ref{fig_DSR_skematic}. 

\begin{figure}[!h]
	\centering
	\includegraphics[scale=0.4]{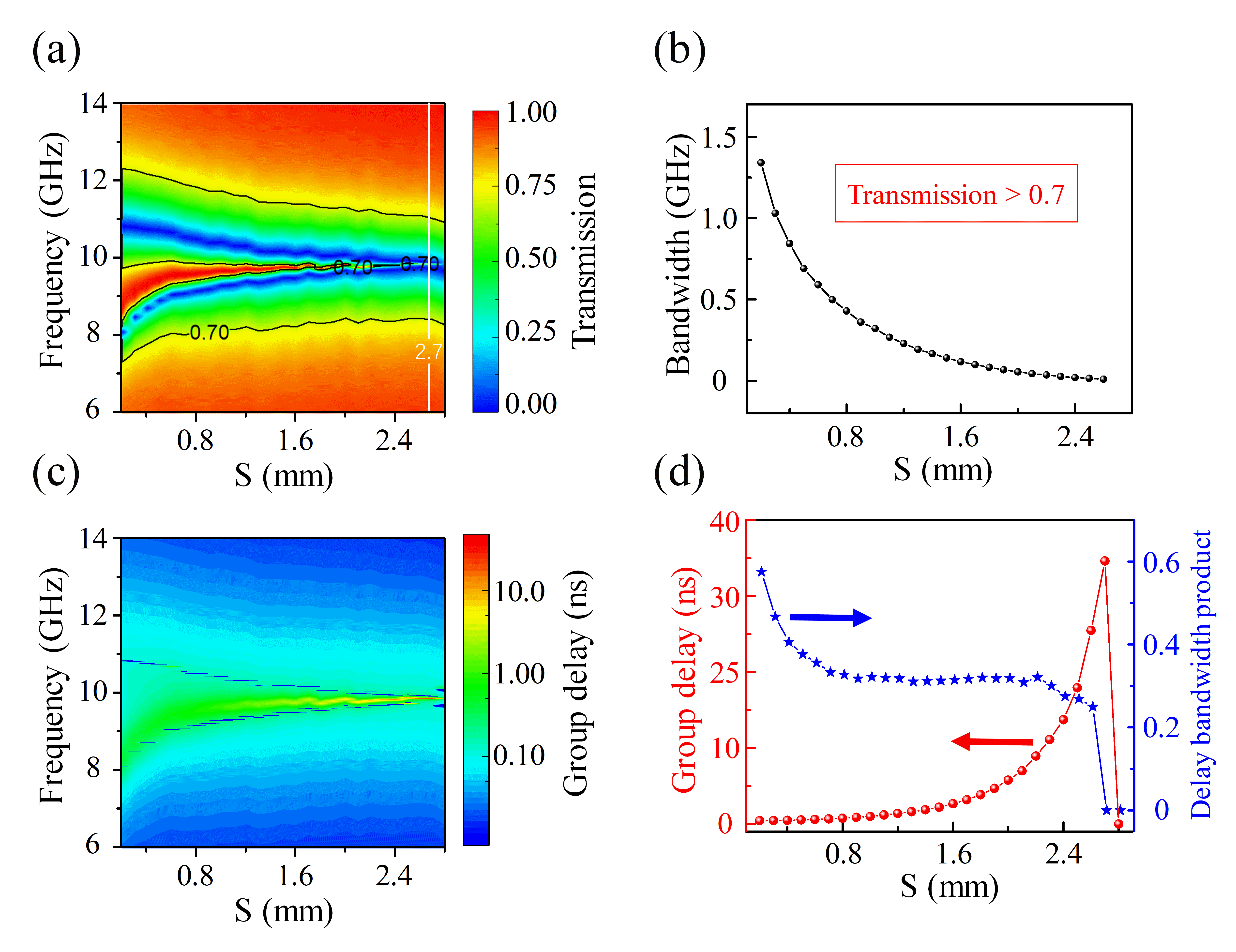}
	\caption{ (a) The simulated 2D map of transmission response spectra as a function of frequency as well as coupling distance ($S$). (b) Transmission bandwidth with amplitude greater than 0.7. (c) 2D map of group delay as a function of frequency as well as coupling distance at. (d) The maximum group delay (red sphere-symbol line) and delay bandwidth product (blue star-symbol line) with the varied $S$. The step of $S$ is 0.1 $mm$}
	\label{fig_DSR_map}
\end{figure}

Meanwhile, we also map the 2D transmission amplitude and phase delay spectra as a function of frequency as well as coupling distance under the condition of stretching loading at the step of 0.1 $mm$, as shown in Fig. \ref{fig_DSR_map}(a) and Fig. \ref{fig_DSR_map}(c). The similar trend is observed with the original design of the transmission bandwidth and group delay. The transmission amplitude is greater than 0.7 when $S$ is below 2.7 $mm$ and the maximum group delay achieves 34.63 ns at the $S$ of 2.7 $mm$, as shown in Fig. \ref{fig_DSR_map}(b) and \ref{fig_DSR_map}(d), respectively. Taking both transmission amplitude and group delay into consideration, the results of DBP with a transmission efficiency greater than 70\% are shown in Fig. \ref{fig_DSR_map}(d) with blue star-symbol line. The maximum DBP is 0.58 at the $S$ of 0.2 $mm$. 

\begin{figure}[!h]
	\centering
	\includegraphics[scale=0.4]{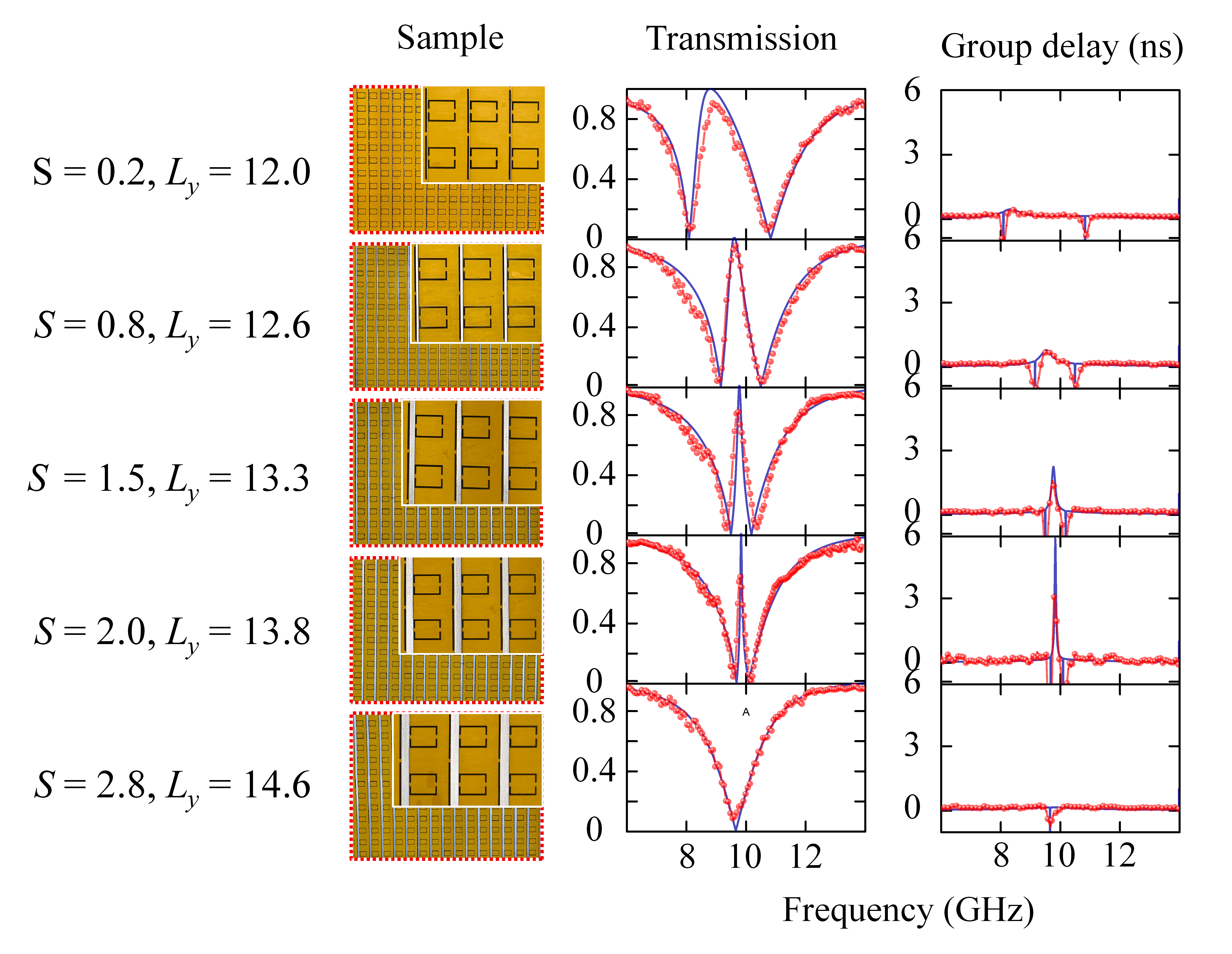}
	\caption{ The optical graphs of the fabricated improved EIT MM with the stretching of $S$ and the simulated (blue solid line) and experimental (red sphere-symbol line) transmission and phase delay results.}
	\label{fig_DSR_test}
\end{figure}

To verify the simulation results, the prototype devices are fabricated using the same method in Fig. \ref{fig_process} and tested with the experimental setup in Fig. \ref{fig_sckematic}(c). The corresponding results are demonstrated in Fig. \ref{fig_DSR_test}. By mechanically stretching, the coupling distance ($S$) can be controlled well as described in the photos. There are also good agreements of the transmission spectra between the simulated (blue solid line) and experimental (red sphere-symbol line) results with the stretching of $S$. Deterioration was observed in measured transmission amplitude and the value of maximum group delay caused by the fabrication tolerances and imperfections of measurement.

Then, compared the proposed structures with other reported EIT MMs within the GHz region in Table \ref{EIT_table_delay}, it can be seen that the DBP and group delay are greatly improved. Specifically, the bandwidth and group delay characteristics of the proposed design can be easily adjusted, which makes great improvements in both dynamic regulation and slow wave characteristics.
\begin{table*}[!h]
	\caption{Comparison with previous studies of EIT MMs in microwave region.}
	\label{EIT_table_delay}
	\centering
	\begin{tabular}{|c|c|c|c|c|c|}
		\hline
		Ref. & Transmission(a.u) & Group delay(ns) & DBP & Polarization dependency & Tunability(Y/N)  \\
		\hline
		\cite{zhang2013large} & 0.78  & 1.46 & - & dependent & N   \\
		\hline
		\cite{li2015electromagnetically} & 0.72 & 2.4 & 0.39 & dependent & N \\
		\hline
		\cite{bagci2018polarization} & 0.70  &  0.333 & 0.45 & independent & N \\
		\hline
		\cite{bagci2019single} & 0.728 & 0.708  & 0.505 & independent  & N \\
		\hline
		\cite{tung2020polarization} & 0.81 & 0.83 & - & independent & N \\
		\hline		
		\cite{yin2020planar} & - & - & 0.528 & dependent & N \\
		\hline
		CW-SRRs &  0.70 & 1.5 & 2.55 & dependent & Y \\
		\hline
		CW-DSR &  0.70 & 0.43  & 0.58 & dependent & Y \\
		\hline
	\end{tabular}
\end{table*}

\section{Conclusion}
In this study, we have discovered a novel way to modulate EIT MMs by analyzing the classical EIT structures, known as CW-SRRs. The proposed mechanism, based on mode coupling, enables reconfigurable bandwidth and group delay of the EIT effect. The underlying mechanism was described through the use of the coupled Lorentz oscillator model and demonstrated through the examination of electric field and surface charge density distributions. The results of our simulations and experiments were in good agreement. Furthermore, drawing inspiration from the tuning mechanism, we presented a block definition design to create stretchable EIT MMs through transfer printing and laser cutting techniques. This approach allowed us to dynamically modulate the bandwidth and group delay of the EIT effect with a large DBP of up to 2.55. In addition, by optimizing the structure design, we proposed CW-DSRs to achieve dynamic modulation of the broadband, narrowband, and suppression of the EIT effect, simply by stretching the device. The ease of fabrication, simplicity of modulation, and compatibility with large-area implementation, make the proposed EIT MM a promising candidate for tunable slow-wave devices and microwave switches.
\section*{Acknowledgment}
This work is supported by the National Natural Science Foundation of China (Nos. 61825102, 52021001, and 61901085), and Medico-Engineering Cooperation Funds from University of Electronic Science and Technology of China (No. ZYGX2021YGLH008)

\bibliography{Refs}

\begin{thebibliography}{10}
\providecommand{\url}[1]{#1}
\csname url@samestyle\endcsname
\providecommand{\newblock}{\relax}
\providecommand{\bibinfo}[2]{#2}
\providecommand{\BIBentrySTDinterwordspacing}{\spaceskip=0pt\relax}
\providecommand{\BIBentryALTinterwordstretchfactor}{4}
\providecommand{\BIBentryALTinterwordspacing}{\spaceskip=\fontdimen2\font plus
\BIBentryALTinterwordstretchfactor\fontdimen3\font minus
  \fontdimen4\font\relax}
\providecommand{\BIBforeignlanguage}[2]{{%
\expandafter\ifx\csname l@#1\endcsname\relax
\typeout{** WARNING: IEEEtran.bst: No hyphenation pattern has been}%
\typeout{** loaded for the language `#1'. Using the pattern for}%
\typeout{** the default language instead.}%
\else
\language=\csname l@#1\endcsname
\fi
#2}}
\providecommand{\BIBdecl}{\relax}
\BIBdecl

\bibitem{doi:10.1063/1.4906109}
\BIBentryALTinterwordspacing
L.~Cong, S.~Tan, R.~Yahiaoui, F.~Yan, W.~Zhang, and R.~Singh, ``Experimental
  demonstration of ultrasensitive sensing with terahertz metamaterial
  absorbers: A comparison with the metasurfaces,'' \emph{Applied Physics
  Letters}, vol. 106, no.~3, p. 031107, 2015. [Online]. Available:
  \url{https://doi.org/10.1063/1.4906109}
\BIBentrySTDinterwordspacing

\bibitem{7393446}
R.~{Yahiaoui}, A.~C. {Strikwerda}, and P.~U. {Jepsen}, ``Terahertz plasmonic
  structure with enhanced sensing capabilities,'' \emph{IEEE Sensors Journal},
  vol.~16, no.~8, pp. 2484--2488, April 2016.

\bibitem{Schurig977}
\BIBentryALTinterwordspacing
D.~Schurig, J.~J. Mock, B.~J. Justice, S.~A. Cummer, J.~B. Pendry, A.~F. Starr,
  and D.~R. Smith, ``Metamaterial electromagnetic cloak at microwave
  frequencies,'' \emph{Science}, vol. 314, no. 5801, pp. 977--980, 2006.
  [Online]. Available: \url{http://science.sciencemag.org/content/314/5801/977}
\BIBentrySTDinterwordspacing

\bibitem{physRevLett.100.207402}
\BIBentryALTinterwordspacing
N.~I. Landy, S.~Sajuyigbe, J.~J. Mock, D.~R. Smith, and W.~J. Padilla,
  ``Perfect metamaterial absorber,'' \emph{Phys. Rev. Lett.}, vol. 100, p.
  207402, May 2008. [Online]. Available:
  \url{https://link.aps.org/doi/10.1103/PhysRevLett.100.207402}
\BIBentrySTDinterwordspacing

\bibitem{doi:10.1002/adma.201200674}
\BIBentryALTinterwordspacing
C.~M. Watts, X.~Liu, and W.~J. Padilla, ``Metamaterial electromagnetic wave
  absorbers,'' \emph{Advanced Materials}, vol.~24, no.~23, pp. OP98--OP120,
  2012. [Online]. Available:
  \url{https://onlinelibrary.wiley.com/doi/abs/10.1002/adma.201200674}
\BIBentrySTDinterwordspacing

\bibitem{Alaee12}
\BIBentryALTinterwordspacing
R.~Alaee, M.~Farhat, C.~Rockstuhl, and F.~Lederer, ``A perfect absorber made of
  a graphene micro-ribbon metamaterial,'' \emph{Opt. Express}, vol.~20, no.~27,
  pp. 28\,017--28\,024, Dec 2012. [Online]. Available:
  \url{http://www.opticsexpress.org/abstract.cfm?URI=oe-20-27-28017}
\BIBentrySTDinterwordspacing

\bibitem{doi:10.1063/1.3608246}
\BIBentryALTinterwordspacing
H.~Li, L.~H. Yuan, B.~Zhou, X.~P. Shen, Q.~Cheng, and T.~J. Cui, ``Ultrathin
  multiband gigahertz metamaterial absorbers,'' \emph{Journal of Applied
  Physics}, vol. 110, no.~1, p. 014909, 2011. [Online]. Available:
  \url{https://doi.org/10.1063/1.3608246}
\BIBentrySTDinterwordspacing

\bibitem{zheng2015metasurface}
G.~Zheng, H.~M{\"u}hlenbernd, M.~Kenney, G.~Li, T.~Zentgraf, and S.~Zhang,
  ``Metasurface holograms reaching 80\% efficiency,'' \emph{Nature
  nanotechnology}, vol.~10, no.~4, p. 308, 2015.

\bibitem{zhao2012twisted}
Y.~Zhao, M.~Belkin, and A.~Al{\`u}, ``Twisted optical metamaterials for
  planarized ultrathin broadband circular polarizers,'' \emph{Nature
  communications}, vol.~3, p. 870, 2012.

\bibitem{PhysRevLett.64.1107}
\BIBentryALTinterwordspacing
S.~E. Harris, J.~E. Field, and A.~Imamo\ifmmode~\breve{g}\else \u{g}\fi{}lu,
  ``Nonlinear optical processes using electromagnetically induced
  transparency,'' \emph{Phys. Rev. Lett.}, vol.~64, pp. 1107--1110, Mar 1990.
  [Online]. Available:
  \url{https://link.aps.org/doi/10.1103/PhysRevLett.64.1107}
\BIBentrySTDinterwordspacing

\bibitem{hau1999light}
L.~V. Hau, S.~E. Harris, Z.~Dutton, and C.~H. Behroozi, ``Light speed reduction
  to 17 metres per second in an ultracold atomic gas,'' \emph{Nature}, vol.
  397, no. 6720, p. 594, Feb 1999.

\bibitem{lukin2001controlling}
M.~Lukin and A.~Imamo{\u{g}}lu, ``Controlling photons using electromagnetically
  induced transparency,'' \emph{Nature}, vol. 413, no. 6853, p. 273, Mar 2001.

\bibitem{doi:10.1002/lpor.201100021}
\BIBentryALTinterwordspacing
I.~Novikova, R.~Walsworth, and Y.~Xiao, ``Electromagnetically induced
  transparency-based slow and stored light in warm atoms,'' \emph{Laser \&
  Photonics Reviews}, vol.~6, no.~3, pp. 333--353, 2012. [Online]. Available:
  \url{https://onlinelibrary.wiley.com/doi/abs/10.1002/lpor.201100021}
\BIBentrySTDinterwordspacing

\bibitem{PhysRevLett.99.123603}
\BIBentryALTinterwordspacing
Y.~Zhang, A.~W. Brown, and M.~Xiao, ``Opening four-wave mixing and six-wave
  mixing channels via dual electromagnetically induced transparency windows,''
  \emph{Phys. Rev. Lett.}, vol.~99, p. 123603, Sep 2007. [Online]. Available:
  \url{https://link.aps.org/doi/10.1103/PhysRevLett.99.123603}
\BIBentrySTDinterwordspacing

\bibitem{Schmidt96}
\BIBentryALTinterwordspacing
H.~Schmidt and A.~Imamoglu, ``Giant kerr nonlinearities obtained by
  electromagnetically induced transparency,'' \emph{Opt. Lett.}, vol.~21,
  no.~23, pp. 1936--1938, Dec 1996. [Online]. Available:
  \url{http://ol.osa.org/abstract.cfm?URI=ol-21-23-1936}
\BIBentrySTDinterwordspacing

\bibitem{yahiaoui2017active}
R.~Yahiaoui, M.~Manjappa, Y.~K. Srivastava, and R.~Singh, ``Active control and
  switching of broadband electromagnetically induced transparency in symmetric
  metadevices,'' \emph{Appl. Phys. Lett.}, vol. 111, no.~2, p. 021101, 2017.

\bibitem{PhysRevLett.101.253903}
\BIBentryALTinterwordspacing
N.~Papasimakis, V.~A. Fedotov, N.~I. Zheludev, and S.~L. Prosvirnin,
  ``Metamaterial analog of electromagnetically induced transparency,''
  \emph{Phys. Rev. Lett.}, vol. 101, p. 253903, Dec 2008. [Online]. Available:
  \url{https://link.aps.org/doi/10.1103/PhysRevLett.101.253903}
\BIBentrySTDinterwordspacing

\bibitem{PhysRevLett.101.047401}
\BIBentryALTinterwordspacing
S.~Zhang, D.~A. Genov, Y.~Wang, M.~Liu, and X.~Zhang, ``Plasmon-induced
  transparency in metamaterials,'' \emph{Phys. Rev. Lett.}, vol. 101, p.
  047401, Jul 2008. [Online]. Available:
  \url{https://link.aps.org/doi/10.1103/PhysRevLett.101.047401}
\BIBentrySTDinterwordspacing

\bibitem{PhysRevB.79.085111}
\BIBentryALTinterwordspacing
R.~Singh, C.~Rockstuhl, F.~Lederer, and W.~Zhang, ``Coupling between a dark and
  a bright eigenmode in a terahertz metamaterial,'' \emph{Phys. Rev. B},
  vol.~79, p. 085111, Feb 2009. [Online]. Available:
  \url{https://link.aps.org/doi/10.1103/PhysRevB.79.085111}
\BIBentrySTDinterwordspacing

\bibitem{PhysRevLett.106.107403}
\BIBentryALTinterwordspacing
C.~Wu, A.~B. Khanikaev, and G.~Shvets, ``Broadband slow light metamaterial
  based on a double-continuum fano resonance,'' \emph{Phys. Rev. Lett.}, vol.
  106, p. 107403, Mar 2011. [Online]. Available:
  \url{https://link.aps.org/doi/10.1103/PhysRevLett.106.107403}
\BIBentrySTDinterwordspacing

\bibitem{liu2009plasmonic}
N.~Liu, L.~Langguth, T.~Weiss, J.~K{\"a}stel, M.~Fleischhauer, T.~Pfau, and
  H.~Giessen, ``Plasmonic analogue of electromagnetically induced transparency
  at the drude damping limit,'' \emph{Nature materials}, vol.~8, no.~9, p. 758,
  2009.

\bibitem{hao2008symmetry}
F.~Hao, Y.~Sonnefraud, P.~V. Dorpe, S.~A. Maier, N.~J. Halas, and
  P.~Nordlander, ``Symmetry breaking in plasmonic nanocavities: subradiant lspr
  sensing and a tunable fano resonance,'' \emph{Nano letters}, vol.~8, no.~11,
  pp. 3983--3988, 2008.

\bibitem{PhysRevB.94.161103}
\BIBentryALTinterwordspacing
M.~Manjappa, Y.~K. Srivastava, and R.~Singh, ``Lattice-induced transparency in
  planar metamaterials,'' \emph{Phys. Rev. B}, vol.~94, p. 161103, Oct 2016.
  [Online]. Available:
  \url{https://link.aps.org/doi/10.1103/PhysRevB.94.161103}
\BIBentrySTDinterwordspacing

\bibitem{xiao2018active}
S.~Xiao, T.~Wang, T.~Liu, X.~Yan, Z.~Li, and C.~Xu, ``Active modulation of
  electromagnetically induced transparency analogue in terahertz hybrid
  metal-graphene metamaterials,'' \emph{Carbon}, vol. 126, pp. 271--278, 2018.

\bibitem{PhysRevA.92.023818}
\BIBentryALTinterwordspacing
J.~A. Souza, L.~Cabral, R.~R. Oliveira, and C.~J. Villas-Boas,
  ``Electromagnetically-induced-transparency-related phenomena and their
  mechanical analogs,'' \emph{Phys. Rev. A}, vol.~92, p. 023818, Aug 2015.
  [Online]. Available:
  \url{https://link.aps.org/doi/10.1103/PhysRevA.92.023818}
\BIBentrySTDinterwordspacing

\bibitem{doi:10.1119/1.1412644}
\BIBentryALTinterwordspacing
C.~L. Garrido~Alzar, M.~A.~G. Martinez, and P.~Nussenzveig, ``Classical analog
  of electromagnetically induced transparency,'' \emph{American Journal of
  Physics}, vol.~70, no.~1, pp. 37--41, 2002. [Online]. Available:
  \url{https://doi.org/10.1119/1.1412644}
\BIBentrySTDinterwordspacing

\bibitem{yan2019terahertz}
X.~Yan, M.~Yang, Z.~Zhang, L.~Liang, D.~Wei, M.~Wang, M.~Zhang, T.~Wang,
  L.~Liu, J.~Xie \emph{et~al.}, ``The terahertz electromagnetically induced
  transparency-like metamaterials for sensitive biosensors in the detection of
  cancer cells,'' \emph{Biosensors and Bioelectronics}, vol. 126, pp. 485--492,
  2019.

\bibitem{yang2019electromagnetically}
M.~Yang, L.~Liang, Z.~Zhang, Y.~Xin, D.~Wei, X.~Song, H.~Zhang, Y.~Lu, M.~Wang,
  M.~Zhang \emph{et~al.}, ``Electromagnetically induced transparency-like
  metamaterials for detection of lung cancer cells,'' \emph{Optics express},
  vol.~27, no.~14, pp. 19\,520--19\,529, 2019.

\bibitem{chen2021analogue}
S.~Chen, T.~Pan, Y.~Peng, G.~Yao, M.~Gao, and Y.~Lin, ``Analogue of
  electromagnetically induced transparency based on bright--bright mode
  coupling between spoof electric localized surface plasmon and electric
  dipole,'' \emph{IEEE Transactions on Microwave Theory and Techniques},
  vol.~69, no.~3, pp. 1538--1546, 2021.

\bibitem{chen2020noncontact}
Z.~Chen, X.~Q. Lin, Y.~H. Yan, F.~Xiao, M.~T. Khan, and S.~Zhang, ``Noncontact
  group-delay-based sensor for metal deformation and crack detection,''
  \emph{IEEE Transactions on Industrial Electronics}, vol.~68, no.~8, pp.
  7613--7619, 2020.

\bibitem{zhang2021highly}
J.~Zhang, N.~Mu, L.~Liu, J.~Xie, H.~Feng, J.~Yao, T.~Chen, and W.~Zhu, ``Highly
  sensitive detection of malignant glioma cells using metamaterial-inspired thz
  biosensor based on electromagnetically induced transparency,''
  \emph{Biosensors and Bioelectronics}, vol. 185, p. 113241, 2021.

\bibitem{ning2019wideband}
R.~Ning, J.~Bao, Y.~Meng, and Z.~Chen, ``Wideband reciprocity tunable
  electromagnetically induced transparency in complementary graphene
  metasurface,'' \emph{Journal of Optics}, vol.~21, no.~4, p. 045106, 2019.

\bibitem{zhang2020active}
C.~Zhang, Y.~Wang, Y.~Yao, L.~Tian, Z.~Geng, Y.~Yang, J.~Jiang, and X.~He,
  ``Active control of electromagnetically induced transparency based on
  terahertz hybrid metal-graphene metamaterials for slow light applications,''
  \emph{Optik}, vol. 200, p. 163398, 2020.

\bibitem{han2020polarization}
L.~Han, Q.~Tan, Y.~Gan, W.~Zhang, and J.~Xiong, ``Polarization-insensitive
  classical electromagnetically induced transparency metamaterial with large
  group delay by dirac semimetal,'' \emph{Results in Physics}, vol.~19, p.
  103377, 2020.

\bibitem{guo2018enhancement}
Z.~Guo, H.~Jiang, Y.~Li, H.~Chen, and G.~Agarwal, ``Enhancement of
  electromagnetically induced transparency in metamaterials using long range
  coupling mediated by a hyperbolic material,'' \emph{Optics express}, vol.~26,
  no.~2, pp. 627--641, 2018.

\bibitem{litt2018hybrid}
D.~B. Litt, M.~R. Jones, M.~Hentschel, Y.~Wang, S.~Yang, H.~D. Ha, X.~Zhang,
  and A.~P. Alivisatos, ``Hybrid lithographic and dna-directed assembly of a
  configurable plasmonic metamaterial that exhibits electromagnetically induced
  transparency,'' \emph{Nano letters}, vol.~18, no.~2, pp. 859--864, 2018.

\bibitem{zhu2022electromagnetically}
L.~Zhu and L.~Dong, ``Electromagnetically induced transparency metamaterials:
  theories, designs and applications,'' \emph{Journal of Physics D: Applied
  Physics}, vol.~55, no.~26, p. 263003, 2022.

\bibitem{pitchappa2016active}
P.~Pitchappa, M.~Manjappa, C.~P. Ho, Y.~Qian, R.~Singh, N.~Singh, and C.~Lee,
  ``Active control of near-field coupling in conductively coupled
  microelectromechanical system metamaterial devices,'' \emph{Applied Physics
  Letters}, vol. 108, no.~11, p. 111102, 2016.

\bibitem{li2015electromagnetically}
H.~Li, S.~Liu, S.~Liu, S.~Wang, H.~Zhang, B.~Bian, and X.~Kong,
  ``Electromagnetically induced transparency with large delay-bandwidth product
  induced by magnetic resonance near field coupling to electric resonance,''
  \emph{Applied Physics Letters}, vol. 106, no.~11, p. 114101, 2015.

\bibitem{su2015broadband}
X.~Su, C.~Ouyang, N.~Xu, S.~Tan, J.~Gu, Z.~Tian, J.~Han, F.~Yan, and W.~Zhang,
  ``Broadband terahertz transparency in a switchable metasurface,'' \emph{IEEE
  Photonics Journal}, vol.~7, no.~1, pp. 1--8, 2015.

\bibitem{wu2011broadband}
C.~Wu, A.~B. Khanikaev, and G.~Shvets, ``Broadband slow light metamaterial
  based on a double-continuum fano resonance,'' \emph{Physical review letters},
  vol. 106, no.~10, p. 107403, 2011.

\bibitem{gu2012active}
J.~Gu, R.~Singh, X.~Liu, X.~Zhang, Y.~Ma, S.~Zhang, S.~A. Maier, Z.~Tian, A.~K.
  Azad, H.-T. Chen \emph{et~al.}, ``Active control of electromagnetically
  induced transparency analogue in terahertz metamaterials,'' \emph{Nature
  communications}, vol.~3, p. 1151, 2012.

\bibitem{chen2008experimental}
H.-T. Chen, J.~F. O'hara, A.~K. Azad, A.~J. Taylor, R.~D. Averitt, D.~B.
  Shrekenhamer, and W.~J. Padilla, ``Experimental demonstration of
  frequency-agile terahertz metamaterials,'' \emph{Nature Photonics}, vol.~2,
  no.~5, p. 295, 2008.

\bibitem{doi:10.1002/adma.201603355}
\BIBentryALTinterwordspacing
M.~Manjappa, Y.~K. Srivastava, L.~Cong, I.~Al-Naib, and R.~Singh, ``Active
  photoswitching of sharp fano resonances in thz metadevices,'' \emph{Advanced
  Materials}, vol.~29, no.~3, p. 1603355, 2017. [Online]. Available:
  \url{https://onlinelibrary.wiley.com/doi/abs/10.1002/adma.201603355}
\BIBentrySTDinterwordspacing

\bibitem{xu2016frequency}
Q.~Xu, X.~Su, C.~Ouyang, N.~Xu, W.~Cao, Y.~Zhang, Q.~Li, C.~Hu, J.~Gu, Z.~Tian
  \emph{et~al.}, ``Frequency-agile electromagnetically induced transparency
  analogue in terahertz metamaterials,'' \emph{Optics letters}, vol.~41,
  no.~19, pp. 4562--4565, 2016.

\bibitem{wang2017hybrid}
S.~Wang, L.~Kang, and D.~H. Werner, ``Hybrid resonators and highly tunable
  terahertz metamaterials enabled by vanadium dioxide (vo 2),''
  \emph{Scientific reports}, vol.~7, no.~1, p. 4326, 2017.

\bibitem{dicken2009frequency}
M.~J. Dicken, K.~Aydin, I.~M. Pryce, L.~A. Sweatlock, E.~M. Boyd,
  S.~Walavalkar, J.~Ma, and H.~A. Atwater, ``Frequency tunable near-infrared
  metamaterials based on vo 2 phase transition,'' \emph{Optics express},
  vol.~17, no.~20, pp. 18\,330--18\,339, 2009.

\bibitem{cao2013plasmon}
W.~Cao, R.~Singh, C.~Zhang, J.~Han, M.~Tonouchi, and W.~Zhang,
  ``Plasmon-induced transparency in metamaterials: Active near field coupling
  between bright superconducting and dark metallic mode resonators,''
  \emph{Applied Physics Letters}, vol. 103, no.~10, p. 101106, 2013.

\bibitem{doi:10.1063/1.4969061}
\BIBentryALTinterwordspacing
P.~Pitchappa, M.~Manjappa, C.~P. Ho, R.~Singh, N.~Singh, and C.~Lee, ``Active
  control of electromagnetically induced transparency with dual dark mode
  excitation pathways using mems based tri-atomic metamolecules,''
  \emph{Applied Physics Letters}, vol. 109, no.~21, p. 211103, 2016. [Online].
  Available: \url{https://doi.org/10.1063/1.4969061}
\BIBentrySTDinterwordspacing

\bibitem{doi:10.1063/1.4943974}
\BIBentryALTinterwordspacing
P.~Pitchappa, M.~Manjappa, C.~P. Ho, Y.~Qian, R.~Singh, N.~Singh, and C.~Lee,
  ``Active control of near-field coupling in conductively coupled
  microelectromechanical system metamaterial devices,'' \emph{Applied Physics
  Letters}, vol. 108, no.~11, p. 111102, 2016. [Online]. Available:
  \url{https://doi.org/10.1063/1.4943974}
\BIBentrySTDinterwordspacing

\bibitem{doi:10.1002/adom.201500676}
\BIBentryALTinterwordspacing
P.~Pitchappa, M.~Manjappa, C.~P. Ho, R.~Singh, N.~Singh, and C.~Lee, ``Active
  control of electromagnetically induced transparency analog in terahertz mems
  metamaterial,'' \emph{Advanced Optical Materials}, vol.~4, no.~4, pp.
  541--547, 2016. [Online]. Available:
  \url{https://onlinelibrary.wiley.com/doi/abs/10.1002/adom.201500676}
\BIBentrySTDinterwordspacing

\bibitem{badloe2021tunable}
T.~Badloe, J.~Lee, J.~Seong, and J.~Rho, ``Tunable metasurfaces: the path to
  fully active nanophotonics,'' \emph{Advanced Photonics Research}, vol.~2,
  no.~9, p. 2000205, 2021.

\bibitem{qi2021stretchable}
D.~Qi, K.~Zhang, G.~Tian, B.~Jiang, and Y.~Huang, ``Stretchable electronics
  based on pdms substrates,'' \emph{Advanced Materials}, vol.~33, no.~6, p.
  2003155, 2021.

\bibitem{li2015low}
H.~Li, S.~Liu, S.~Liu, S.~Wang, G.~Ding, H.~Yang, Z.~Yu, and H.~Zhang,
  ``Low-loss metamaterial electromagnetically induced transparency based on
  electric toroidal dipolar response,'' \emph{Applied Physics Letters}, vol.
  106, no.~8, p. 083511, 2015.

\bibitem{tassin2012electromagnetically}
P.~Tassin, L.~Zhang, R.~Zhao, A.~Jain, T.~Koschny, and C.~M. Soukoulis,
  ``Electromagnetically induced transparency and absorption in metamaterials:
  the radiating two-oscillator model and its experimental confirmation,''
  \emph{Physical review letters}, vol. 109, no.~18, p. 187401, 2012.

\bibitem{han2016tunable}
S.~Han, L.~Cong, H.~Lin, B.~Xiao, H.~Yang, and R.~Singh, ``Tunable
  electromagnetically induced transparency in coupled three-dimensional
  split-ring-resonator metamaterials,'' \emph{Scientific reports}, vol.~6, p.
  20801, 2016.

\bibitem{tian2017low}
Y.~Tian, S.~Hu, X.~Huang, Z.~Yu, H.~Lin, and H.~Yang, ``Low-loss planar
  metamaterials electromagnetically induced transparency for sensitive
  refractive index sensing,'' \emph{Journal of Physics D: Applied Physics},
  vol.~50, no.~40, p. 405105, 2017.

\bibitem{PhysRevLett.102.053901}
\BIBentryALTinterwordspacing
P.~Tassin, L.~Zhang, T.~Koschny, E.~N. Economou, and C.~M. Soukoulis,
  ``Low-loss metamaterials based on classical electromagnetically induced
  transparency,'' \emph{Phys. Rev. Lett.}, vol. 102, p. 053901, Feb 2009.
  [Online]. Available:
  \url{https://link.aps.org/doi/10.1103/PhysRevLett.102.053901}
\BIBentrySTDinterwordspacing

\bibitem{tsakmakidis2017breaking}
K.~Tsakmakidis, L.~Shen, S.~Schulz, X.~Zheng, J.~Upham, X.~Deng, H.~Altug,
  A.~Vakakis, and R.~Boyd, ``Breaking lorentz reciprocity to overcome the
  time-bandwidth limit in physics and engineering,'' \emph{Science}, vol. 356,
  no. 6344, pp. 1260--1264, 2017.

\bibitem{zhao2018maximization}
Z.~Zhao, Y.~Chen, Z.~Gu, and W.~Shi, ``Maximization of terahertz slow light by
  tuning the spoof localized surface plasmon induced transparency,''
  \emph{Optical Materials Express}, vol.~8, no.~8, pp. 2345--2354, 2018.

\bibitem{zhang2013large}
F.~Zhang, X.~He, X.~Zhou, Y.~Zhou, S.~An, G.~Yu, and L.~Pang, ``Large group
  index induced by asymmetric split ring resonator dimer,'' \emph{Applied
  Physics Letters}, vol. 103, no.~22, p. 221904, 2013.

\bibitem{bagci2018polarization}
F.~Bagci and B.~Akaoglu, ``A polarization independent electromagnetically
  induced transparency-like metamaterial with large group delay and
  delay-bandwidth product,'' \emph{Journal of Applied Physics}, vol. 123,
  no.~17, p. 173101, 2018.

\bibitem{bagci2019single}
------, ``Single and multi-band electromagnetically induced transparency-like
  effects with a four-fold symmetric metamaterial design,'' \emph{Materials
  Research Express}, vol.~6, no.~5, p. 055806, 2019.

\bibitem{tung2020polarization}
B.~S. Tung, B.~X. Khuyen, P.~T. Linh, N.~T. Tung, D.~H. Manh, and V.~D. Lam,
  ``Polarization-insensitive electromagnetically-induced transparency in planar
  metamaterial based on coupling of ring and zigzag spiral resonators,''
  \emph{Modern Physics Letters B}, vol.~34, no.~10, p. 2050093, 2020.

\bibitem{yin2020planar}
X.~Yin, M.~Wu, Y.~Liu, and C.~Huang, ``A planar metamaterial based on metallic
  rectangular-ring pair for narrow electromagnetically induced
  transparency-like effect,'' \emph{Journal of Applied Physics}, vol. 128,
  no.~6, p. 065105, 2020.

\end{thebibliography}
\bibliographystyle{IEEEtran}
\end{document}